\documentstyle[12pt]{article}

\newcommand{\ue}%
{\mbox{c\hspace{-0.4em}\rule[
0.5ex]{0.3em}{0.04ex}\hspace{0.1em}}}
\newcommand{\Ue}%
{\mbox{C\hspace{-0.6em}\rule[
0.75ex]{0.4em}{0.04ex}\hspace{0.2em}}}
\newlength{\signlength}%
\newcommand{\gs}%
{\settowidth{\signlength}{$<$}%
\raisebox{-0.35\signlength}%
{\makebox[1.7\signlength][c]{\makebox[0.pt]{$\sim$}%
\raisebox{0.6\signlength}{\makebox[0.pt]{$>$}}}}}
\newcommand{\ls}%
{\settowidth{\signlength}{$<$}%
\raisebox{-0.35\signlength}%
{\makebox[1.7\signlength][c]{\makebox[0.pt]{$\sim$}%
\raisebox{0.6\signlength}{\makebox[0.pt]{$<$}}}}}
\renewcommand{\ge}%
{\settowidth{\signlength}{$>$}%
\setlength{\unitlength}{0.1\signlength}%
\mbox{\begin{picture}(17,7.5)%
\linethickness{0.045\signlength}%
\put(0.0,1){\makebox(17,7.5){$>$}}%
\multiput(4.5,-1)(0.25,0.125){32}{\line(1,0){0.25}}%
\end{picture}}}
\renewcommand{\le}%
{\settowidth{\signlength}{$>$}%
\setlength{\unitlength}{0.1\signlength}%
\mbox{\begin{picture}(17,7.5)%
\linethickness{0.045\signlength}%
\put(0.0,1){\makebox(17,7.5){$<$}}%
\multiput(12.,-1)(-0.25,0.125){32}{\line(-1,0){0.25}}%
\end{picture}}}
\title{}
\author{}
\date{}
\begin{document}
\maketitle
\thispagestyle{empty}

\newpage
\begin{center}

Magnetic ordering in layered high temperature superconductors

G.G.Sergeeva

\end{center}

Kharkov Institute of Physics and Technology, Academicheskaya 1,
310108, Kharkov,

Ukraine

\begin{abstract}
It is discussed the scenario of two-step magnetic ordering
in layered HTS after charge ordering. At the temperature decreasing
the transition from 3D Heisenberg spin behavior to 2D XY coupling
of the Cu spins occurs in di-stripes. Further temperature decreasing
leads to the spin glass transition at $T_g$.

\end{abstract}

$keyword$

charge ordering, magnetic ordering  \hspace{10mm}
hole concentration in $Cu O_2$-planes

2D XY-model, quasi 2D XY-model  \hspace{10mm}
3D-spin glass transition

\vspace{10mm}
1.INTRODUCTION
\vspace{05mm}

The conclusion about the decisive role of fluctuational
antiferromagnetic (AFM)
excitations with 3D Heisenberg spin behavior
in HTS has been substantiated reliably to now.
The opening of a "spin pseudogap" on the Fermi surface
was connected at first with strong AFM correlations. But later
the propositions that the formation of the pseudogap
as well as superconductivity are associated
with charge ordering (dynamic analog of phase separation)
 were discussed (see reviews [1,2])  and were
experimentally confirmed [3,4].
Simple example of charge ordering was observed
in Bi2212, where
in the Cu-O plane metallic (met-) stripes with orthorombic structure
 and the stripes of the dielectric (di-) tetrogonal phase with a short
range AFM correlations were found out at $T<T_{ch}$
($T_{ch}$ is the temperature of charge ordering) [5].
In spite of essential anisotropy
of exchange constants of in-plane ($J_0$) and enter-plane ($J_1$)
interactions, $ J_0/J_1>10^{3}$, strong AFM
fluctuations in layered HTS prevent to 2D Heisenberg ordering
in di-stripes. In this paper it is discussed the scenario of
two-step magnetic ordering in layered HTS which is
based on the known properties of 2D XY-model [6-7].  It is
 shown, that after charge ordering in Cu-O planes the
transition from 3D Heisenberg spin behavior to 2D XY coupling
of the Cu spins occurs
at $ T=T_{BKT} =T_{sp}< T_{ch}$ in the di-stripes, where
$T_{BKT}$ is Berezinskii-Kosterlitz-Thouless temperature .
Further temperature decreasing leads to the spin
glass transition at $T_g$,
which was predicted for quasi-2D XY-model in Ref.8.

\vspace{10mm}
2. RESULTS.
\vspace{05mm}

 It is known that in 2D XY-model the
phase transition occurs
which leads to the formation of phase
with power law decreasing of  in-plane
correlations, and to the peculiarities
of temperature dependencies of the
resistivity, magnetic susceptibility
and specific heat   at $T<T_{sp}$ [6].
For layered  systems the order
parameter of quasi-2D XY-model $q=0$ at  $T >T_{sp}$ and
 $q\sim J_1 ^{\Delta/(2-\Delta)}$ at $T<T_{sp}$, where $\Delta$
 is scaling dimensionality [7].

If $p_{cr}$ is hole concentration in $Cu O_2$-planes, at which the
compound becomes the superconductor, and if
the effective hole concentration in the di-stripes
$p_{sh}^{\ast}< p_{cr}$,
the temperature of spin ordering, $ T_{sp}$, may be large enough,
$T_{sp}(p_{sh}) \sim T_N(p_{sh}^{\ast})$ (here
$p_{sh}$ is the hole concentration in $Cu O_2$-planes
and $T_N$ is the Neel temperature).

Taking into  account that exchange constant  $J_1 \ll J_0$, as
well as that 2D XY spin ordering  at $T_{sp}$ occurs independently
in each di-stripe, order parameter $q\ll 1$ and the sample
  remains non-magnetized. But after influence of magnetic field
  weak summarized magnetization can be observed.

The temperature decreasing can lead to the
 3D-spin glass transition at $T_g\ll T_{sp}$,
if each di-stripe is considered as  2D XY-model with
 occasional anisotropy and if the coupling between layers
has special form  $ J_1 cos n (\Theta_{i+1}- \Theta _i)$
( $\Theta$ is the angle which is determined the direction of
spin in $i$-plane, $n$ is the anisotropy order). Such transition
for quasi-2D XY-model was predicted for quasi-2D XY-model [8] and
for HTS was observed at $ \mu sR$ measurements [4].

\vspace{10mm}
3.DISCUSSION.
\vspace{05mm}

For magnetic compounds with high anisotropy of exchange constants
the two step spin ordering is well known and the peculiarities of
temperature dependencies of the resistivity, magnetic susceptibility
and specific heat at temperature of 2D XY ordering were observed.
The first indirect evidence that spins may cross over to
2D XY-like behavior at $T\sim 20 $ K, was observed  in neutron
measurements of single crystal $La_{2-x} Sr_x Cu O_4$
which is in the intermediate region of $x=0.04$
with neither long range AFM order nor superconductivity [9].
There are enough experimental evidences of two step magnetic
ordering in compounds with $p_{sh}< p_{cr}$
obtained in neutron and magnetic
measurements, and $ \mu sR$ studies ( see Ref.10 and references
therein). Magnetic phase diagram as
a function of $p_{sh}$ for $La_{2-x} Sr_x Cu O_4$
and $Y_{1-x} Ca_x Ba_2 Cu_3 O_{6.02}$ ($0<x<0.11$) was
obtained in $ \mu sR$ measurements [4].
The coexistence of spin glass state and
superconducting state for hole concentration
$0.06<p_{sh}<0.10$ was found out.
The observations of the oscillations of attenuation decrement
after influence of magnetic field and residual
magnetization at 300 K in Bi2223 ceramics with 30 \% Bi2212 are
indirect evidences of 2D XY magnetic ordering
and weak summarized magnetization of di-stripes
(see Ref.11 and references therein).
It should be interesting to carry out the
measurements of  $T_{sp}$ for $p_{sh}> p_{cr}$
 such as were performed for $p_{sh}< p_{cr}$.
These measurements of temperature of 2D XY
magnetic ordering are seemed to very important
and experimentally feasible step in studies of nature of HTS.

\vspace{10mm}
References

1. V.Barzykin and D.Pines, Phys.Rev. B 52, 13585 (1995).

2. S.G.Ovchinnikov, Usp.Fiz.Nauk 167, 1042 (1997);
G.G.Sergeeva et al.Fiz.Nizk.Temp.

24, 1029 (1998)[Low Temp.phys. 24, 771 (1998)]

3. A.A.Zakharov et al. Physica C 223, 157 (1994)

4. Ch.Niedermayer et al. Phys.Rev.Lett. 80, 3843 (1998)

5. A.Bianconi and M.Missori, J.Phys.I (Paris),4, 361 (1994)

6. V.L.Berezinskii, JETP, 59, 907 (1970); ibid. 61, 1144 (1971).
J.M. Kosterlitz,

D.J.Thouless. J.Phys., C6, 1181 (1973)

7. V.L.Berezinskii, A.Ya.Blank, JETP,64, 723 (1973).

8. Vik.Dotsenko and M.V.Feigelman, JETP, 83, 345 (1982).

9. B.Keimer et al. Phys.Rev. B 46, 14034 (1992)

10. F.C.Chou et al. Phys.Rev.Lett. 75, 2204 (1995)

11. B.G.Lazarev et al. Fiz.Nizk.Temp.22, 819 (1996) [Low Temp.Phys 22,
4629 (1996)];

\end{document}